%% file: main.tex
\documentclass[amsthm,thmtools]{ifacconf}
\usepackage{natbib}
\usepackage{amsmath,amssymb,amsfonts,graphicx,url}

\usepackage{times}
\usepackage{dsfont}
\usepackage{mathtools}
\usepackage[utf8]{inputenc}
\usepackage[T1]{fontenc}

\usepackage{graphicx,color,fancyhdr}
\usepackage{subfigure}
\usepackage{enumerate}
\usepackage{multirow}

\newcommand{\indep}{\perp \!\!\! \perp}

\input{_definitions}

\renewcommand{\Re}{{\mathbb R}}
\newcommand\numberthis{\addtocounter{equation}{1}\tag{\theequation}}
\usepackage{comment}

\begin{document}

\begin{frontmatter}
	
    \title{Causal Optimal Coupling for Gaussian Input-Output Distributional Data} 
	
	\author[XDR]{Daran Xu} 
	\author[AHT]{Amirhossein Taghvaei} 
	\address[XDR]{Department of Applied Mathematics, University of Washington, Seattle,
		WA (e-mail: daranxu@uw.edu).}
	
	\address[AHT]{Department of Aeronautics and Astronautics, University of Washington, Seattle, 
		WA (e-mail: amirtag@uw.edu).}

\begin{abstract}
We study the problem of identifying an optimal coupling between input–output distributional data generated by a causal dynamical system. The coupling is required to satisfy prescribed marginal distributions and a causality constraint reflecting the temporal structure of the system. We formulate this problem as a Schrödinger Bridge, which seeks the coupling closest—in Kullback–Leibler divergence—to a given prior while enforcing both marginal and causality constraints. 
For the case of Gaussian marginals and general time-dependent quadratic cost functions, we derive a fully tractable characterization of the Sinkhorn iterations that converges to the optimal solution.  Beyond its theoretical contribution, the proposed framework provides a principled foundation for applying causal optimal transport methods to system identification from distributional data.
\end{abstract}
	
	
\end{frontmatter}

\section{Introduction}
Consider an unknown input-output system. 
Suppose we observe independent realizations of the input and output processes, but the underlying pairing between the observations is unknown. Our objective is to identify an input--output relationship that is consistent with the observed data while remaining close to a given reference model. Formally, let ${U}:=\{U_t\}_{t=1}^T \sim \mu$ and ${Y}:=\{Y_t\}_{t=1}^T \sim \nu$ be two stochastic processes with known probability distributions $\mu$ and $\nu$, respectively.  The joint probability law of $(U,Y)$ is unknown. To infer the input-output relationship, we search over the set of all couplings between $\mu$ and $\nu$, where a causal constraint is imposed on these couplings to account for the temporal structure of the input-output relationship. In particular, a coupling $\pi$ of $(U,Y)$ satisfies the  causality constraint if 
\begin{equation}
    \label{eq:causal-constraint}
    (Y_1, \dots, Y_t)  \indep_{(U_1,\dots,U_t)}(U_{t+1}, \dots, U_T )
\end{equation}
for all $t=1,2,\ldots, T-1$; that is to say, given the past inputs, the future inputs are independent of the past measurements. 
We use  $ \Pi(\mu, \nu) $ to denote the collection of all couplings with marginals $\mu$ and $\nu$, and we use  $ \Pi_c(\mu, \nu) $ to denote the couplings in $ \Pi(\mu, \nu) $ that satisfy the causality constraint. 

{\sloppy
Inspired by the classic Schr\"odinger Bridge problem, we formulate the system identification as the optimization problem:
\begin{equation}
    \min_{\pi \in  \Pi_c(\mu,\nu)}\,D(\pi , \gamma)
    \label{problem: SB} 
\end{equation}
}
where $D_{}(\pi, \gamma):=\int \log(\frac{\ud \pi}{\ud \gamma}) \ud \pi$ denotes the Kullback-Leibler (KL) divergence between the coupling $\pi$ and a reference coupling $\gamma$. The reference model $\gamma$ encodes prior structural or modeling assumptions about the input--output system. By  decomposing the KL-divergence as
\[
D(\pi,\gamma) = \int \log (\frac{\ud (\mu \otimes \nu)}{\ud \gamma}) \, \mathrm{d}\pi + D(\pi,\mu \otimes \nu),
\]
and 
defining the cost function $c := \log (\frac{\ud (\mu \otimes \nu)}{\ud \gamma})$, the  optimization 
problem~\eqref{problem: SB} can be equivalently written as
\begin{equation}
\label{eq:problem-formulation-2}
\min_{\pi \in \Pi_c(\mu,\nu)} \; \int c \, \mathrm{d}\pi + D(\pi,\mu \otimes \nu) .
\end{equation}
This formulation recovers the classical entropy-regularized optimal coupling problem, with the key distinction that admissible couplings are required to satisfy a causality constraint. We refer to it as the entropic causal optimal transport (ECOT) problem. 

In this paper, we focus on the case where the input and output are one-dimensional, and the reference coupling $\gamma$ is induced by a linear Gaussian model. For the simplest reference model,
\[
Y_t = U_t + \text{``Gaussian noise''},
\]
{The resulting cost is equivalent to $\sum_{t=1}^T (y_t - u_t)^2$, rescaled by the variance of Gaussian noise, which coincides with the cost underlying the Wasserstein--$2$ distance between $\mu$ and $\nu$ up to a constant factor.} More generally, when $\gamma$ corresponds to an arbitrary linear Gaussian input--output model  of the form~\eqref{eqn:dynamic}, the cost takes the quadratic form
\begin{equation}
c({u}_{1:T},{y}_{1:T}) =\sum_{t=1}^{T}\frac{1}{2\epsilon_t^2} (y_t - {\mathbf  h}_t^\top  u_{1:t}  -  {\mathbf  f}_t^\top{y}_{1:t-1}-b_t)^2,
\label{eq:general-quad-cost}
\end{equation}
where, for $t=1,2,\ldots,T$, ${u}_{1:t}=(u_1,u_2,\ldots,u_t)$, ${y}_{1:t}=(y_1,y_2,\ldots,y_t)$, and ${\mathbf  f}_t, {\mathbf  h}_t,b_t,\epsilon_t$ depend on the reference model (see Section~\ref{sec:problem} for details).  Our goal is to solve the ECOT problem \eqref{eq:problem-formulation-2} for this general class of quadratic costs.

Causal optimal transport has been studied in \cite{veraguas2017causaltransportdiscretetime}, where the causality structure is exploited to derive a dynamic programming formulation of the optimal coupling problem. However, this approach is restricted to settings in which the input process $\{U_t\}$ is Markovian and the cost function admits the separable form $\sum_{t=1}^T c_t(u_t,y_{1:t}$), which significantly limits its applicability and does not contain the cost~\eqref{eq:general-quad-cost} that we are interested in. 
The ECOT problem is investigated in \cite{eckstein2023computationalmethodsadaptedoptimal}, where abstract Sinkhorn iterations are proposed to approximate the optimal regularized coupling. Our work builds on this paper with the goal of making these Sinkhorn iterations explicit, tractable, and implementable for the case of Gaussian marginals and the general quadratic cost in \eqref{eq:general-quad-cost}.
For Gaussian marginals and the squared loss $\sum_{t=1}^T \|y_t - u_t\|^2$, closed-form solutions are available under a \emph{bicausal} constraint, in which causality is imposed in both directions—from $U$ to $Y$ and from $Y$ to $U$—~\cite{gunasingam2025adaptedoptimaltransportgaussian,acciaio2025entropicadaptedwassersteindistance}. 
In contrast to these works, we consider a more general quadratic cost of the form \eqref{eq:general-quad-cost} and impose only a \emph{one-sided} causality constraint. This modeling choice is motivated by the input--output interpretation of the data and allows for richer temporal dependencies while remaining computationally tractable.

In addition to the causal optimal transport literature, our work builds on a line of research focused on identifying dynamical models from distributional or aggregate data~\cite{bernstein2016consistently,chen2021optimal, haasler2021control, mascherpa2023estimating, terpin2024dynamic, emerick2024causal,mascherpa2025convex}.

Here is an outline  of the paper:
\begin{itemize}
    \item In Section~\ref{sec:problem}, we present the causal linear Gaussian input-output reference model, and connect it to the proposed quadratic cost function. 
    \item In Section~\ref{section: sinkhorn solving COT}, we provide explicit Sinkhorn iterations to solve the proposed ECOT problem, for the case with Gaussian marginal probability measures and quadratic cost. 
    \item In Section~\ref{sec:example}, we provide a numerical example that demonstrates the optimal causal coupling obtained through Sinkhorn iterations.  
\end{itemize}

Theoretical analysis of the results and the details of the proposed numerical algorithm are omitted on the account of space and would appear as part of the long version of this paper.

\section{Problem Formulation}\label{sec:problem}
Our goal is to solve the  ECOT problem~\eqref{eq:problem-formulation-2}.  
We focus on the reference coupling $\gamma$ induced by a linear Gaussian causal input-output model of the form  
\begin{equation}
    Y_t = {\mathbf h_t}^{\top} U_{1:t}+ {\mathbf f_t}^{\top}   Y_{1:t-1}+b_t +w_t , \label{eqn:dynamic}
\end{equation}
for $t\in \{1,\cdots, T\}$,  
where $U_{1:t}=(U_1,\ldots,U_t)$ and $  Y_{1:t-1}^{}=(Y_1,\ldots,Y_{t-1})$ are the history of input and output up to time $t$ and $t-1$, respectively, and ${\mathbf h_t}\in \mathbb{R}^{t}$, ${\mathbf f_t}\in \mathbb{R}^{t-1}$, and $b_t\in\mathbb{R}$  are the time-dependent model parameters.  The noise terms $w_t \sim \mathcal{N}(0,\epsilon_t^2)$ are assumed to be independent of each other, and of $\{U_{1:t}, Y_{1:t-1}\}$. 
Note that the selected input-output model~\eqref{eqn:dynamic} is inherently non-anticipating, ensuring that the resulting reference $\gamma$ satisfies the causal constraint.
With this choice for the model, the conditional probability density of $Y_t=y_t$, given $ U_{1:t}=u_{1:t}$ and $Y_{1:t-1}=y_{1:t-1}$, is  equal to
\begin{equation*}
    \frac{1}{\sqrt{2\pi} \epsilon_t}\exp\left(-\frac{1}{2\epsilon_t^2} \left(y_t - \mathbf{h}_t^\top u_{1:t}  -  \mathbf{f}_t^\top{y}_{1:t-1}-b_t\right)^2\right). 
\end{equation*}
concluding the cost function~\eqref{eq:general-quad-cost} after removing terms that only depend on the marginal distributions and do not affect the optimal coupling. Compared with the classical cost $\sum_{t=1}^T (u_t-y_t)^2$, the cost in \eqref{eq:general-quad-cost}  is more general, allowing the interactions over time,  capturing temporal dependencies within the dynamic.

\begin{remark}The general reference  model \eqref{eqn:dynamic}  covers the special case of the linear Gaussian state-space model with input $U_t$ and output $Y_t$: 
    \begin{align*}
    X_{t}&=F X_{t-1}+B U_{t}+Q^{\frac{1}{2}} V_{t} \\
    Y_t &=H X_t+R^{\frac{1}{2}} W_t \numberthis \label{eqn: kamlan filter}
\end{align*}
where $V_{t} $ and ${W_t}$ are independent and identically distributed Gaussian random variables. Under this model, the conditional density of $Y_t$ given $ U_{1:t}$ and $ Y_{1:t-1}$ is Gaussian with the mean and variance determined by the Kalman filter. In particular, the mean would be an affine function of the past inputs $U_{1:t}$ and observations  $Y_{1:t-1}$, and the errors are independent and Gaussian, which concludes an instant of the reference model~\eqref{eqn:dynamic}. 
\end{remark}

\section{Sinkhorn Iteration in the Gaussian Setting}
\label{section: sinkhorn solving COT}
In this section, we introduce the Sinkhorn iterations for solving the ECOT problem \eqref{eq:problem-formulation-2}. Following the abstract framework developed in~\cite{eckstein2023computationalmethodsadaptedoptimal}, we derive an explicit and implementable formulation of the iterations tailored to the problem setting described in Section~\ref{sec:problem}.

The Sinkhorn algorithm is formulated as an iterative two-step procedure consisting of alternating projections onto the marginal constraint sets. The iteration is initialized at the reference (prior) coupling, $\pi^{ \{0\}}=\gamma$. For $k=0,1,2,\ldots$, the coupling is updated via the following two optimization steps:
\begin{align}
	\text{(odd iteration)}\quad &\pi^{ \{2k+ 1 \} } :=  \argmin_{\pi \in\Pi_c(\mu, *)} D_{}(\pi, \pi^{ \{2k\} }) \label{eqn: odd sinkhorn}, \\
		\text{(even iteration)}\quad &\pi^{ \{ 2k+2 \} } := \argmin_{\pi \in \Pi(*, \nu)} D_{}(\pi, \pi^{ \{2k+ 1\} }) \label{eqn: even sinkhorn},
\end{align}
where $\Pi_c(\mu, *)$ denotes the set of all causal couplings of $(U,Y)$ with the fixed $U$-marginal equal to $\mu$, and $\Pi(*,\nu)$ denotes the set of all  couplings with the fixed $Y$-marginal equal to $\nu$. The odd and even steps can therefore be interpreted as projections onto $\Pi_c(\mu, *)$ and $\Pi(*,\nu)$, respectively. The existence and uniqueness of the minimizers in these projection steps are established in \cite[Lem. 6.2 and 6.4]{eckstein2023computationalmethodsadaptedoptimal} and \cite[Thm. 1.10]{nutz2021introductionentropic}, using both primal and dual characterizations.

Within this abstract Sinkhorn framework, our objective is to derive explicit formulas for the odd and even iterations in the case where the marginals $\mu$ and $\nu$, as well as the reference coupling $\gamma$, are Gaussian.  We first observe that the Gaussian family is invariant under both projection steps; consequently, $\pi^{\{k\}}$ is  Gaussian for $k=0,1,2.\ldots$. This structural property allows the iterations to be implemented either via the primal or the dual formulation, leading to two algorithmic realizations. For brevity, we present the primal formulation.

The following notation is needed to state our result. The marginal Gaussian distribution $\mu$ for $U$ is $\mathcal N(m^\mu,\Sigma^\mu)$ with the mean vector $m^\mu \in \Re^T$ and the covariance matrix $\Sigma^\mu \in \Re^{T\times T}$. It is convenient to express $\mu$ via its temporal disintegration:
\begin{align*}
	\mu(u_{1:T}) = \prod_{t=1}^{T}\mu(u_t\mid u_{1:t-1})
\end{align*}
where $\mu(u_t|u_{1:t-1})$ denotes the conditional density of $U_t$ given $U_{1:t-1}$. Each conditional distribution has the Gaussian form
$$\mathcal N(K^{\mu}_{u_t|u_{1:t-1}}u_{1:t-1}+
m^{\mu}_{u_t|u_{1:t-1}},\Sigma^{\mu}_{u_t|u_{1:t-1}})$$  with the  parameters $$(K^{\mu}_{u_t|u_{1:t-1}},m^{\mu}_{u_t|u_{1:t-1}},\Sigma^{\mu}_{u_t|u_{1:t-1}})\in \Re^{t-1}\times \Re \times \Re,$$for $t=1,\ldots,T$, uniquely determined by $(m^\mu,\Sigma^\mu)$. These conditional parameters provide an equivalent parametrization of the Gaussian measure $\mu$, which will be used to express the odd step of the Sinkhorn iteration.

Similarly, we introduce notation for a joint Gaussian measure $\pi$ on $(U,Y)$ and its disintegration. 
We write
\[
\pi = \mathcal N\!\left(
\begin{bmatrix}
m^{\pi}_u \\[2pt]
m^{\pi}_y
\end{bmatrix},
\begin{bmatrix}
\Sigma^{\pi}_u & \Sigma^{\pi}_{u,y} \\
\Sigma^{\pi}_{y,u} & \Sigma^{\pi}_y
\end{bmatrix}
\right).
\]
Its temporal factorization (disintegration) is given by
\[
\pi(u_{1:T},y_{1:T})
= \prod_{t=1}^{T}
\pi(u_t \mid u_{1:t-1}, y_{1:t-1})
\pi(y_t \mid u_{1:t}, y_{1:t-1}).
\]
For notational convenience, define 
\[
z_{1:t} := (u_{1:t}, y_{1:t}), 
\qquad
\tilde z_{1:t} := (u_{1:t+1}, y_{1:t}).
\]
Under Gaussianity, the conditional distributions $\pi(u_t \mid z_{1:t-1})$ and $\pi(y_t \mid \tilde z_{1:t-1})$ admit the representations
\begin{align*}
&\scalebox{0.8}{$
\mathcal N\!\left(
K^{\pi}_{u_t|u_{1:t-1}} u_{1:t-1}
\!+\!
K^{\pi}_{u_t|y_{1:t-1}} y_{1:t-1}
\!+\!
m^{\pi}_{u_t|z_{1:t-1}},
\Sigma^{\pi}_{u_t|z_{1:t-1}}
\right)$}, \\
&\scalebox{0.8}{$
\mathcal N\!\left(
K^{\pi}_{y_t|u_{1:t}} u_{1:t}
\!+\!
K^{\pi}_{y_t|y_{1:t-1}} y_{1:t-1}
\!+\!
m^{\pi}_{y_t|\tilde z_{1:t-1}},
\Sigma^{\pi}_{y_t|\tilde z_{1:t-1}}
\right)$},
\end{align*}
respectively. 
The collection of parameters
\begin{align*}
\big(
K^{\pi}_{u_t|u_{1:t-1}},
K^{\pi}_{u_t|y_{1:t-1}},
m^{\pi}_{u_t|z_{1:t-1}},
\Sigma^{\pi}_{u_t|z_{1:t-1}}
\big), \\
\big(
K^{\pi}_{y_t|u_{1:t}},
K^{\pi}_{y_t|y_{1:t-1}},
m^{\pi}_{y_t|\tilde z_{1:t-1}},
\Sigma^{\pi}_{y_t|\tilde z_{1:t-1}}
\big),
\end{align*}
for $t=1,\ldots,T$, provides an equivalent parametrization of the joint Gaussian measure $\pi$. 
We adopt this parametrization to characterize the optimal coupling in the odd step of the Sinkhorn iteration.

Finally,  the conditional distribution $\pi(z_{t:T}|z_{1:t-1})$ is Gaussian and can be represented as   
\begin{align*}
    & { \mathcal N( K^{\pi}_{z_{t:T}|z_{1:t-1}} z_{1:t-1} \!+\!   m^{\pi}_{z_{t:T}|z_{1:t-1}} ,\Sigma^{\pi}_{z_{t:T}|z_{1:t-1}}).}
\end{align*}
For later use, we decompose the linear term in the conditional mean as
\begin{align*}
		K^{\pi}_{z_{t:T}|z_{1:t-1}}&z_{1:t-1} = 		K^{\pi}_{z_{t:T}|u_{1:t-1}} u_{1:t-1}   \\&+ 		K^{\pi}_{z_{t:T}|y_{1:t-2}} y_{1:t-2} + 	K^{\pi}_{z_{t:T}|y_{t-1}}y_{t-1}. 
\end{align*}

With these notations, we are ready to present the solution to the odd iteration of the Sinkhorn algorithm. 
\begin{theorem}[Odd iteration]\label{theorem: primal odd}
	If $\mu$ is a non-degenerate Gaussian distribution, 
	and $\inf_{\pi' \in \Pi_c(\mu,*)}\,D_\text{KL}(\pi',\gamma) < \infty$, 
	the solution $\pi$ to the following causal projection problem
	\begin{align*}
		\pi =\argmin_{\pi' \in \Pi_c(\mu,*)}\,D_\text{KL}(\pi',\gamma),	\end{align*}
	is also Gaussian. The parameters of the conditional distribution $\pi(u_t|u_{1:t-1},y_{1:t-1})$ are given by
	\begin{align*}
		K^{\pi}_{u_t|u_{1:t-1}}  & =
		K^{\mu}_{u_t|u_{1:t-1}} , \quad 
		K^{\pi}_{u_t|y_{1:t-1}}  = 0, \\
		m^{\pi}_{u_t|u_{1:t-1}}  & = 	m^{\mu}_{u_t|u_{1:t-1}} , \quad 
		\Sigma^{\pi}_{u_t|u_{1:t-1}}  =  \Sigma^{\mu}_{u_t|u_{1:t-1}}. 
	\end{align*}
	 The parameters of the conditional distribution $\pi(y_t|u_{1:t},y_{1:t-1})$ are given by
	\begin{align*}
		K^{\pi}_{y_t|y_{1:t-1}} &= \alpha_t	K^{\gamma}_{y_t|y_{1:t-1}}   + (1-\alpha_t) K^{\psi}_{y_t|y_{1:t-1}} ,  \\
		K^{\pi}_{y_t|u_{1:t}}  &= \alpha_t	K^{\gamma}_{y_t|u{1:t}}   + (1-\alpha_t) K^{\psi}_{y_t|u{1:t}} ,  \\
		m^{\pi}_{y_t|\tilde z_{1:t-1}}   & =  \alpha_t	m^{\gamma}_{y_t|\tilde z_{1:t-1}}  + (1-\alpha_t) m^{\psi}_{y_t|\tilde z_{1:t-1}} , \\
		\Sigma^{\pi}_{y_t|\tilde z_{1:t-1}}  &= \alpha_t \Sigma^{\gamma}_{y_t|\tilde z_{1:t-1}} ,
	\end{align*}
 where
\begin{align*}
	\alpha_t &= \frac{	\Sigma^{\psi}_{y_t|\tilde z_{1:t-1}} }{	\Sigma^{\gamma}_{y_t|\tilde z_{1:t-1}} + 	\Sigma^{\psi}_{y_t|\tilde z_{1:t-1}} }, \\
	\Sigma^{\psi}_{y_t|\tilde z_{1:t-1}}   & = \frac{1}{\delta_t^\top (\Sigma^{\gamma}_{z_{t+1:T}|z_{1:t}})^{-1} \delta_t}, \\
	K^{\psi}_{y_t|y_{1:t-1}} &=   -	\Sigma^{\psi}_{y_t|\tilde z_{1:t-1}} \delta_t^\top (\Sigma^{\gamma}_{z_{t+1:T}|z_{1:t}})^{-1} \Delta^y_t, \\
	K^{\psi}_{y_t|u_{1:t}} &=   -	\Sigma^{\psi}_{y_t|\tilde z_{1:t-1}}  \delta_t^\top (\Sigma^{\gamma}_{z_{t+1:T}|z_{1:t}})^{-1} \Delta^u_t, \\
    m^{\psi}_{y_t|\tilde z_{1:t-1}}&= -	\Sigma^{\psi}_{y_t|\tilde z_{1:t-1}} \delta_t^\top (\Sigma^{\gamma}_{z_{t+1:T}|z_{1:t}})^{-1}  \Delta m_t,\\
	\delta_t& = K^{\pi}_{z_{t+1:T}|y_t}- K^{\gamma}_{z_{t+1:T}|y_t}, \\		
	\Delta^y_t & = K^{\pi}_{z_{t+1:T}|y_{1:t-1}}- K^{\gamma}_{z_{t+1:T}|y_{1:t-1}},\\
	\Delta^u_t & = K^{\pi}_{z_{t+1:T}|u_{1:t}}- K^{\gamma}_{z_{t+1:T}|u_{1:t}},\\
    \Delta m_t&= m^{\pi}_{z_{t:T}|z_{1:t-1}}-m^{\gamma}_{z_{t:T}|z_{1:t-1}}.
\end{align*}
	\end{theorem}

Next, we are going to present the result for the even iteration. Similar to $\mu$, 
the marginal Gaussian distribution $\nu$ for $Y$ is represented as $\mathcal N(m^\nu,\Sigma^\nu)$ with the mean vector $m^\nu \in \Re^T$ and the covariance matrix $\Sigma^\nu \in \Re^{T\times T}$. In the even iteration of the Sinkhorn algorithm, we do not perform disintegration over time. Rather, for a joint Gaussian probability measure $\pi$ on $(U,Y)$, we now consider its disintegration of $U$ given $Y$
$$\pi(u_{1:T},y_{1:T})=\pi(u_{1:T}|y_{1:T}) \pi(y_{1:T})$$
where the conditional distribution $\pi(u_{1:T}|y_{1:T})$ is 
$$\mathcal{N}(K^{\pi}_{u|y} y_{1:T}+m^{\pi}_{u|y},\Sigma^{\pi}_{u|y} ),$$
and $\pi(y_{1:T})$ is a Gaussian distribution
$\mathcal{N}(m^{\pi}_y,\Sigma^{\pi}_y)$. 
The set of parameters 
$$(K^{\pi}_{u|y} ,m^{\pi}_{u|y},\Sigma^{\pi}_{u|y} ),\quad (m^{\pi}_y,\Sigma^{\pi}_y)$$
represent an equivalent parametrization of the
joint Gaussian measure $\pi$. We use this equivalent parametrization to identify the optimal coupling in the even iteration of the Sinkhorn algorithm.
\begin{theorem}[Even iteration]\label{theorem: primal even} 
If $\nu$ is a non-degenerate Gaussian distribution, 
    and $\inf_{\pi' \in \Pi(*,\nu)}\,D_\text{KL}(\pi',\gamma) < \infty$, 
    the solution $\pi$ to the following projection problem
	\begin{align*}
		\pi =\argmin_{\pi' \in \Pi(*,
        \nu)}\,D_\text{KL}(\pi',\gamma)	\end{align*}
 is also Gaussian, with the parameters given by:
\begin{align*}
&K^{\pi}_{u|y}=K^{\gamma}_{u|y},\quad 
 m^{\pi}_{u|y} =m^{\gamma}_{u|y},\quad 
 \Sigma^{\pi}_{u|y} =\Sigma^{\gamma}_{u|y} \\
 &m^{\pi}_{y} =m^{\gamma}_{},\quad 
 \Sigma^{\pi}_{y} =\Sigma^{\gamma}_{}.
\end{align*}
 
\end{theorem}

After running the odd iteration \eqref{eqn: odd sinkhorn} and the even iteration \eqref{eqn: even sinkhorn} of the Sinkhorn algorithm until convergence, we obtain a joint Gaussian coupling $\pi^{*}$ 
on $(U,Y)$. From this coupling, we can recover a linear Gaussian input-output model of the form~\eqref{eqn:dynamic} by 
conditioning of $Y_t$ on $\text{past}_t=\{U_{1:t},Y_{1:t-1}\}$:
\begin{align*}
&[\mathbf{h}_t^\top, \mathbf{f}_t^\top] 
= \text{Cov}^{\pi^*}({Y_t,\text{past}_t}) 
  {\text{Cov}^{\pi^*}({\text{past}_t, \text{past}_t})}^{-1}, \numberthis \label{eqn: conditional formula to get dynamic from the coupling} \\
&b_t = \mathbb{E}^{\pi^*}[Y_t] - \left( \mathbf{h}_t^\top \, \mathbb{E}^{\pi^*}[U_{1:t}] + \mathbf{f}_t^\top \, \mathbb{E}^{\pi^*}[Y_{1:t-1}] \right), \\
&\epsilon_t^2=\\
&\scalebox{0.8}{$ \text{Cov}^{\pi^*}({Y_t,Y_t}) 
-\text{Cov}^{\pi^*}({Y_t,\text{past}_t} )
  {\text{Cov}^{\pi^*}({\text{past}_t, \text{past}_t})}^{-1} 
\text{Cov}^{\pi^*}({\text{past}_t,Y_t})$}, 
\end{align*}
where we use the notation $\text{Cov}^\pi({u,v})$ for the covariance matrix  between $(u,v)$ under the probability measure $\pi$. 

\section{Numerical Example}
\label{sec:example}

We consider the time interval $[0,1]$ discretized into $T=128$ steps. 
The marginal processes are specified as stationary Gaussian processes with
\begin{align*}
\mathbb{E}^{\mu}[U_t] &= 1, 
& \mathrm{Cov}^{\mu}(U_t,U_{t+s}) &= K(s;1), \\
\mathbb{E}^{\nu}[Y_t] &= 0, 
& \mathrm{Cov}^{\nu}(Y_t,Y_{t+s}) &= K(s;0.5),
\end{align*}
where
\[
K(s;\sigma) = \exp\!\left(-\frac{|s|}{2\sigma^2}\right).
\]
As a reference model, we use \eqref{eqn: kamlan filter} with parameters 
$F=B=Q=H=R=1$.

Under this setup, Problem~\ref{problem: SB} is solved using the Sinkhorn scheme described in Section~\ref{section: sinkhorn solving COT}. 
We alternately apply the odd \eqref{eqn: odd sinkhorn} and even \eqref{eqn: even sinkhorn} projections until convergence, defined by
$
\|\pi^{\{k+1\}} - \pi^{\{k\}}\| < 10^{-6}
$.
For comparison, we also solve the entropic OT problem over the enlarged feasible set $\Pi(\mu,\nu)$, without the causality constraint. 
This corresponds to the Sinkhorn iteration~\eqref{eqn: odd sinkhorn}-\eqref{eqn: even sinkhorn}, with the projection set changed from $\Pi_c(\mu,*)$ to $\Pi(\mu,*)$. 

\begin{figure}[t]
    \centering    \includegraphics[width=0.4\textwidth]{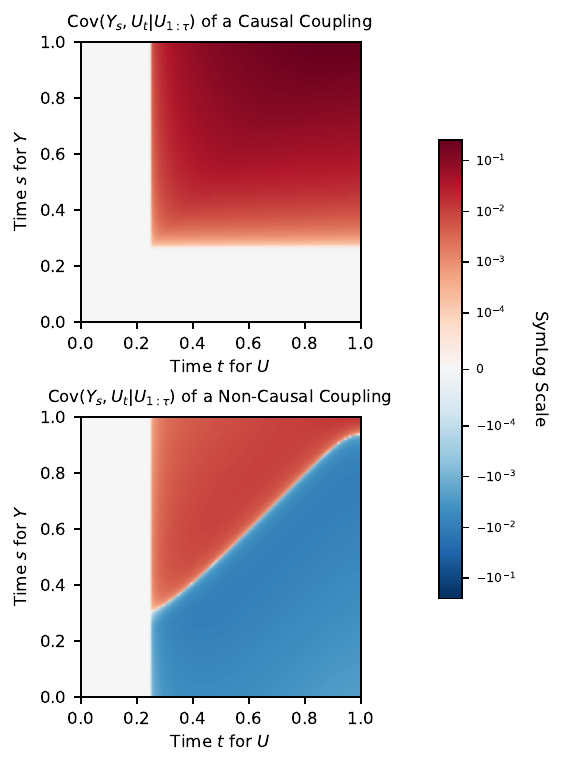}
    \caption{
$\text{Cov}(U_t,Y_s|U_{1:\tau})$ with $\tau=0.25$. The top panel is a causal case, and the bottom panel is a non-causal case. The magnitude is shown on a symmetric logarithmic scale, preserving both sign and zero.}    \label{fig:cross_cov_result}
\end{figure}

Figure~\ref{fig:cross_cov_result} shows the heatmap of the conditional cross-covariance 
\[
\mathrm{Cov}(U_t, Y_s \mid U_{1:\tau}),
\qquad \tau = 0.25,
\]
under both the causal and non-causal couplings. For $t<\tau$, the conditional covariance is zero in both cases, since conditioning fixes the variables. 
For $t>\tau$ and $s<\tau$, the causal coupling yields zero conditional covariance, reflecting the non-anticipative constraint. 
In contrast, the non-causal coupling exhibits nonzero values in this region, indicating dependence on past outputs. When $t>\tau$ and $s>\tau$, the causal coupling produces a consistent, strictly positive conditional covariance structure. 
The non-causal solution, however, displays sign changes across the diagonal and anti-symmetric patterns, revealing anticipatory correlations that violate causality.

\section{Conclusion}\label{sec:conclusion}
 This paper formulates system identification from distributional input–output data as an entropic causal optimal transport (ECOT) problem and solves it via Sinkhorn iterations. We focus on the linear–Gaussian setting, where both input and output marginals are Gaussian, and the reference model is linear Gaussian. In this case, we derive an explicit and computationally tractable form of the Sinkhorn iterations, yielding a fully implementable algorithm. These results illustrate the effectiveness of causal optimal transport for identifying dynamical relationships directly at the level of probability laws.

Future research directions include extending the framework to nonlinear reference models and non-Gaussian marginals, with the goal of making the ECOT formulation applicable to generative modeling of time-series data.

\bibliography{ifacconf} 
\end{document}

%% file: _definitions.tex
\def\Re{\mathbb{R}}

\def\argmin{\mathop{\text{\rm arg\,min}}}

\def\notes#1{\marginpar{\tiny #1}\typeout{Notes!
Notes!
Notes!
}}
\renewcommand{\notes}[1]{\typeout{notes!}}

\def\Re{\field{R}}

\def\S{O}

\newtheorem{theorem}{Theorem}

\newtheorem{remark}{Remark}
\newtheorem{proposition}{Proposition}

\def\beq{\begin{eqnarray}} 
\def\bc{\begin{center}} 
\def\be{\begin{enumerate}}
\def\bi{\begin{itemize}} 
\def\bs{\begin{small}}
\def\bS{\begin{slide}}
\def\ec{\end{center}} 
\def\ee{\end{enumerate}}
\def\ei{\end{itemize}}
\def\es{\end{small}}
\def\eS{\end{slide}}
\def\eeq{\end{eqnarray}}

\newcommand{\newP}[1]{\noindent{\bf #1:}}

\newcommand{\ud}{\,\mathrm{d}}

\def\Re{\mathbb{R}}

\def\argmin{\mathop{\text{\rm arg\,min}}}

\renewcommand{\Re}{\mathbb{R}}

\newcounter{rmnum}

\newcounter{anum}

